\documentclass[twocolumn]{aastex701}
\usepackage[vlined]{algorithm2e}
\usepackage{xspace}
\usepackage{xcolor}

\newcommand{\assign}[1]{{#1} \gets }
\newcommand{\assigntwo}[1]{\makebox[0.3cm][l]{$#1$} \gets }
\newcommand{\WHFastFiveTwelve}{\texttt{\mbox{WHFast512}}\xspace}
\newcommand{\WHFast}{\texttt{\mbox{WHFast}}\xspace}
\newcommand{\REBOUND}{\texttt{\mbox{REBOUND}}\xspace}
\makeatletter
\renewcommand{\algocf@captiontext}[2]{\vspace{8pt}\small\rmfamily {\AlCapNameFnt{#1.} #2\endgraf}\vspace{5pt}}%
\makeatother

\begin{document}

\title{A Very Fast Symplectic Integrator for Planetary Systems: WHFast512 in x86 Assembly}

\author[orcid=0000-0002-7622-3222]{Rishit Dagli}
\email{rishit@cs.toronto.edu}
\affiliation{Dept. of Computer Science, University of Toronto, Toronto, 40 St. George Street, Toronto, ON M5S 2E4, Canada}

\author[orcid=0000-0003-1927-731X]{Hanno Rein}
\email{hanno.rein@utoronto.ca}
\affiliation{Dept. of Computer Science, University of Toronto, Toronto, 40 St. George Street, Toronto, ON M5S 2E4, Canada}
\affiliation{Dept. of Physical and Environmental Sciences, University of Toronto at Scarborough, Toronto, Ontario, M1C 1A4, Canada}
\affiliation{Dept. of Physics, University of Toronto, Toronto, Ontario, M5S 3H4, Canada}
\affiliation{Dept. of Astronomy and Astrophysics, University of Toronto, Toronto, Ontario, M5S 3H4, Canada}

\begin{abstract}
    Long-term integrations of planetary systems are used in a variety of astrophysical applications, for example to determine a system's stability.
    In this paper, we describe significant speed and accuracy improvements to the \WHFastFiveTwelve integrator.
    By completely rewriting \WHFastFiveTwelve in x86 assembly, we gain fine-grained control over CPU instructions and are able to keep the entire simulation state in CPU registers, only writing to memory when an output is required.
    We also reduce the number of branches in the code, improve the Kepler solver's speed and accuracy with a convergence check, switch the Hamiltonian splitting to Jacobi coordinates, implement symplectic correctors, and add support for ejection and collision detection.
    As a benchmark we run simulations of the Solar System with all eight planets and general relativistic corrections for 5~Gyr.
    With our improvements, we can integrate this system in 9~hours, more than $8\times$ faster than the standard version of \WHFast.
    This makes \WHFastFiveTwelve by far the fastest N-body integrator for this kind of simulation.
    We run extensive tests to verify the integrator's accuracy, with a particular focus on removing any long-term bias, and show that the accuracy is now five orders of magnitude better for typical setups compared to the earlier version of \WHFastFiveTwelve.
    The new \WHFastFiveTwelve integrator is freely available in the \REBOUND integrator package.
\end{abstract}


\section{Introduction} 
Symplectic integrators that make use of the Wisdom-Holman splitting \citep{Wisdom1981, WisdomHolman1991, Kinoshita1991} are widely used to integrate particles in systems where the motion is dominated by one central object such as planetary systems.
Many different variants of the Wisdom-Holman (WH) integrator exist \citep[see e.g.][]{LaskarRobutel2001,HernandezDehnen2016, ReinTamayoBrown2019}.

In contrast to many other problems in computational astrophysics, small-$N$ simulations are difficult to parallelize efficiently due to their inherently sequential nature.
Each timestep itself can be calculated easily, but a single simulation might require more than $10^{12}$ timesteps which need to be performed one after another because the result of each timestep is required for the next one\footnote{\cite{Saha1997} use a time parallel method with some success. The method achieves a speed-up of $50\times$, but it comes with a huge cost: $1000\times$ the computational resources.}.
As such, even on modern CPUs and with access to large computer clusters, an individual N-body simulation can take days, weeks, or even months to run.

\begin{figure*}[t]
\includegraphics[width=\textwidth]{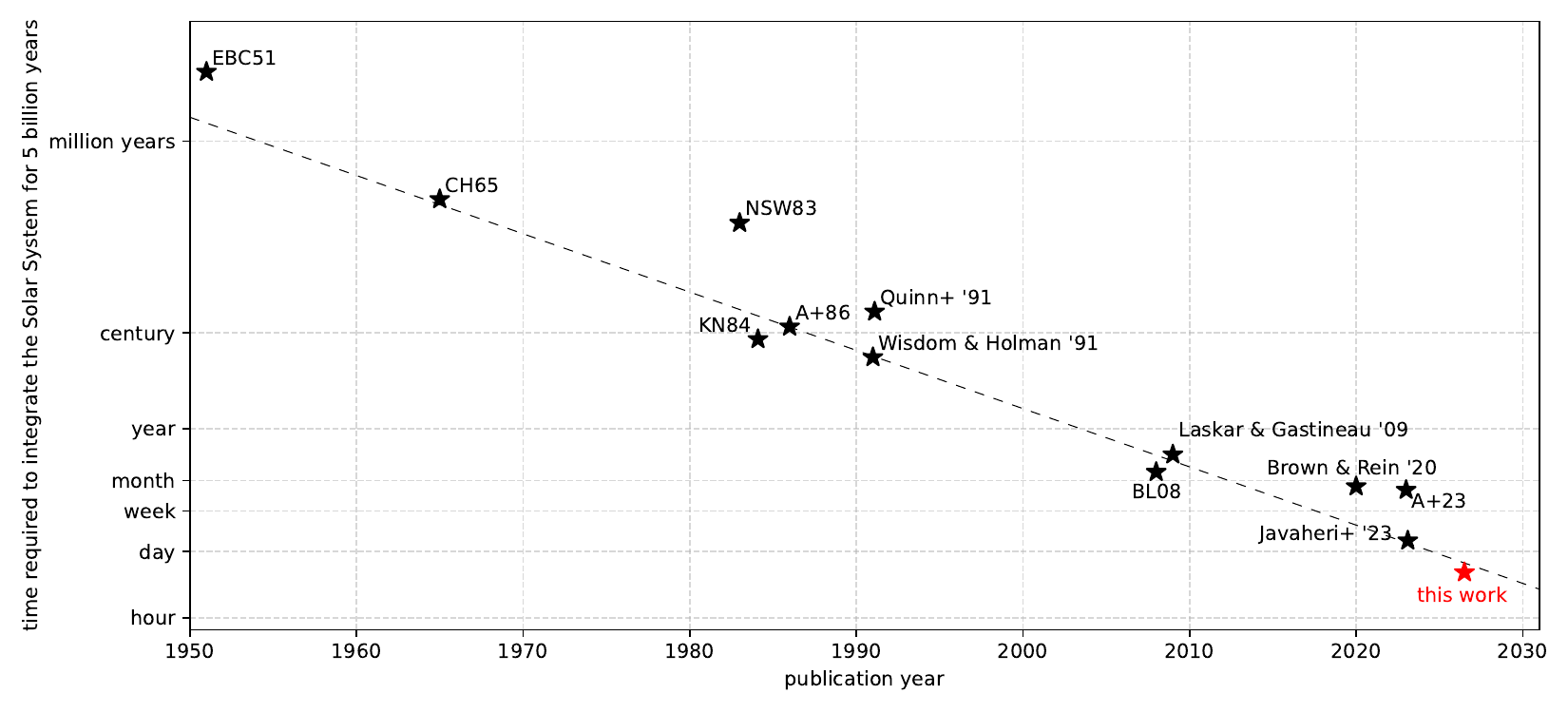} 
\caption{
    The wall-clock time required for an integration of the Solar System with all eight planets for 5~Gyr as reported by \cite{Eckert1951,Cohen1965,Newhall1983, Kinoshita1984, Applegate1986, Quinn1991, WisdomHolman1991, Batygin2008, LaskarGastineau2009, Abbot2023} and \cite{Javaheri2023}.
    The bottom right data point shows the result reported in this paper.
    The four rightmost data points all use a version of either \WHFast or \WHFastFiveTwelve.
    The dashed line shows an exponential function, halving the walltime every 2.5~yrs.
    The data has been adopted from Sam Hadden, \url{https://shadden.github.io/nbody_history/}.
    A historical N-body scaling figure for large $N$ appears in \cite{DehnenRead2011}.
    \label{fig:timerequiredfor5gyr}}
\end{figure*}

Over the years, improvements in computer performance, particularly CPU speed, have significantly reduced the runtime of astrophysical simulations.
In Fig.~\ref{fig:timerequiredfor5gyr} we plot the wall-clock time (elapsed real time) required to integrate the Solar System with all eight planets for 5~Gyr, as reported in 13~publications spanning eight decades.
To account for differences in timestep and particle number, the reported simulation efficiency has to be normalized. 
If $t_{\rm wall}$ is the reported wall-clock time for a simulation integrating $N$ planets for a time $t_{\rm sim}$, then we define $t_{\rm wall, norm}=t_{\rm wall} \cdot (5\,{\rm Gyr} / t_{\rm sim} )\cdot (P_{\min} / 88\,{\rm d})\cdot (8/N)^2$, where $P_{\min}$ is the shortest orbital period of the simulated planets.
As indicated by the dashed line, the wall-clock time falls off exponentially, halving on average every 2.5~yr.
This is consistent with Moore's law \citep{Moore1965}, which predicts that the number of transistors on a computer chip doubles approximately every 2~yr \citep{Moore1975}.

What Fig.~\ref{fig:timerequiredfor5gyr} hides is that CPU clock rates have barely increased since approximately the year~2000 \citep{Danowitz2012}.
This is due to physical constraints, such as CPU heat dissipation, and architectural choices that favor parallel processing of large data sets, including multi-core CPUs and GPUs \citep{Esmaeilzadeh2011}.
The performance gains shown in Fig.~\ref{fig:timerequiredfor5gyr} over the last two decades, particularly those reported by \cite{Javaheri2023} and the present work, are not due to higher clock rates but rather due to improvements in algorithms and the use of advanced CPU instructions.

AVX512 describes a set of CPU features which extends the standard x86~instruction set by adding new instructions that operate on 512-bit wide registers.
Such a register can store up to 8~double precision floating point numbers.
One AVX512 instruction can then operate on all 8~floating point numbers in parallel.
This makes it ideal for scientific computing where many floating point operations are required.
Parallelization with AVX512 has the big advantage that all operations happen on one CPU, thus avoiding the need for communication between different CPU/GPU cores or cluster nodes. 
This is particularly important for the kind of astrophysical simulation considered here because one timestep can take as little as~$100{\rm ns}$\footnote{This corresponds to a light travel time of only $30{\rm m}$, implying that no computer larger than $30{\rm m}$ can possibly be used to speed up the calculation of one timestep. This is a strict causal limit and does not depend on the specific technology used.}.
Some compilers can automatically generate AVX512 instructions, but as we show in this paper, their capabilities are nowhere near what can be achieve by assembling the instructions by hand.
The fact that the Solar System has eight~planets is a fortunate coincidence as it allows us to fully occupy a register with exactly one floating point number for each of the eight~planets.
This allows us to speed up simulations of the Solar System particularly well.
However, \WHFastFiveTwelve can be used to integrator other planetary systems as long as the number of planet is less or equal to eight.

Our work builds on that described in \cite{Javaheri2023}, which uses explicit AVX512 instructions for N-body simulations of planetary systems.
This first version of the \WHFastFiveTwelve integrator was written in C99 using intrinsic functions for the AVX512 instructions.
Although this approach is convenient from a programming perspective, it leaves many decisions about register use and the order of operations to the compiler.
The present paper describes several significant improvements to the \WHFastFiveTwelve integrator, which we refer to as \WHFastFiveTwelve Version 2.
Specifically, we reduce the wall-clock time by a factor of~4 while increasing the accuracy by 5~orders of magnitude.

We note that \cite{Nitadori2006} and \cite{Tanikawa2012} also use SSE and AVX instructions to speed up N-body simulations.
Their methods use a Hermite scheme and target star clusters with more particles than the systems considered here, $N\sim10^3-10^5$.
Like us, they accelerate their algorithms by writing crucial parts in assembly.
Work by \cite{DehnenHernandez2017} also describes some success in using vectorization to speed up large simulations with~$N=1024$.

The plan for this paper is as follows.
In Section~\ref{sec:algo} we describe the algorithmic improvements that we made such as switching to Jacobi coordinates and changes to the Kepler solver and the Stiefel functions.
We then focus on the performance optimizations in Section~\ref{sec:optimizations} where we discuss our choice to reimplement \WHFastFiveTwelve in assembly, the advantages of keeping the entire simulation in registers, an efficient matrix multiplication algorithm, how we avoid branching, and when the use of approximate operations is acceptable. 
Various tests are presented in Section~\ref{sec:tests} which demonstrate the accuracy and performance of our new integrator.
We discuss in Section~\ref{sec:discussion} why we think our integrator is almost optimal, making it very hard to achieve further speed improvements.
We conclude in Section~\ref{sec:conclusions}.

\section{Algorithmic Improvements}\label{sec:algo}
The main idea of a WH integrator is to split the motion of each planet into two parts: motion around the central object on a Keplerian orbit and perturbations from interactions with the other planets.
This perturbative approach reduces the error, relative to a non-perturbative integrator, by a factor comparable to the Jupiter-to-Sun mass ratio, roughly~$10^{-3}$ in the Solar System.

To perform this splitting, a WH integrator must solve two problems in a leap-frog fashion.
First, planets are moved along Keplerian orbits in the \textit{Kepler step}.
Kepler's equation has no closed form solution, so this step is performed iteratively.
Second, forces from planet-planet interactions are applied in the \textit{interaction step}.
This step can be computed explicitly in any inertial frame but can be computationally expensive because it involves $N\!\cdot\!(N\!-\!1)/2$ force evaluations.
For a more comprehensive introduction to the WH integrator see the original papers by \cite{Wisdom1981} and \cite{WisdomHolman1991} and specifically for the \WHFast and previous \WHFastFiveTwelve implementations of the WH integrator, see \cite{ReinTamayo2015} and \cite{Javaheri2023} respectively.

Parallelizing the Kepler and interaction steps poses different challenges and we describe our algorithmic improvements in this section.

\subsection{Jacobi Coordinates} \label{sec:jacobi}
The separation into Keplerian motion and a perturbation part is not unique and can be done in different ways, each of which has its own advantages and disadvantages. 
\cite{ReinTamayo2019} lists the two Hamiltonian splittings most relevant to this paper: DHC and Jacobi coordinates.
Other choices exist, including canonical heliocentric coordinates \citep{LaskarRobutel1995}, inertial Cartesian coordinates, barycentric coordinates, and the maps of \cite{HernandezBertschinger2015} as well as \cite{Hernandez2016}.
The first version of \WHFastFiveTwelve used democratic heliocentric coordinates (DHCs) for the Hamiltonian splitting.
This was done because it was easier: DHCs do not require coordinate transformations between the interaction and Kepler steps.
Furthermore, planets do not need to be ordered.
However, for the same timestep, a Hamiltonian splitting based on DHCs is less accurate than one based on Jacobi coordinates \citep{Farres2013}.
In particular, DHCs are not well suited for long-term integrations of planetary systems, as described in detail in \cite{Rein2026}.
This is because the DHC splitting of the Hamiltonian introduces an additional numerical precession that can easily dominate over physical effects.
To avoid unphysical results, one needs to reduce the timestep by almost an order of magnitude.
However, doing so eliminates the potential speed-up gained by \WHFastFiveTwelve.

For this reason, we use Jacobi coordinates in the new version of \WHFastFiveTwelve.
The main downside of Jacobi coordinates is that they require coordinate transformations between each Kepler step and interaction step.
Specifically, one needs to transform the positions from Jacobi coordinates to heliocentric coordinates before the interaction step and transform the accelerations from heliocentric coordinates to Jacobi coordinates afterward.

One key insight that allows us to implement these coordinate transformations efficiently is that they can be written as linear transformations, specifically $8\times8$~triangular matrices for an eight-planet system.
The matrix coefficients are constants and can be calculated once at the beginning of the simulation because they depend only on the particle masses, not on positions or velocities \citep[see, e.g.,][]{Beust2003}.
Our specific implementation of an efficient matrix-vector multiplication is described in Section~\ref{sec:matrix}.

Hamiltonian splittings with either Jacobi or DHCs admit symplectic correctors, which can reduce errors by orders of magnitude with almost zero overhead \citep{Wisdom1996,HernandezDehnen2016}.
However, with general relativistic corrections enabled (see next section), the \cite{Wisdom1996} style correctors in DHCs no longer work.
This is another reason for using Jacobi coordinates.
We implement symplectic correctors of order~17 using a concatenation of Kepler and interaction steps.

\subsection{General Relativity, Close Encounters, and Ejections}\label{sec:gr}
The stability of planetary systems, particularly the Solar System, is sensitive to the precise precession frequencies of the planets \citep{Laskar2008, LaskarGastineau2009, BoueLaskarFarago2012, BrownRein2023}.
We therefore implement an option to include a general relativistic correction to Newtonian gravity in the form of an additional $1/r^2$ potential centred on the central object.
This first order correction mimics the effects of general relativistic precession \citep{Nobili1986}.
Because we use Jacobi coordinates, we already calculate the heliocentric distance of each planet in the interaction step.
Thus, adding this correction term has minimal cost because no additional square roots or divisions are required.
We note that this treatment of GR is only an approximation.
It is well suited for cases like the Solar System where accurate secular precession rates are important.
However, it introduces small errors in the orbital frequencies and might thus not be sufficient for other cases, in particular in strong gravity regimes.

We also implement optional checks for close encounters and ejections at every timestep.
The ejection check compares the heliocentric distance of each planet with a preset value.
The simulation stops when the distance exceeds this value.
Similarly, the encounter check compares the distance between each pair of planets with a preset value.
The simulation stops when the distance falls below this value.
Once again, we can reuse distances that are already calculated in the interaction step, so these checks do not add significant overhead.

All three optional features increase the runtime by at most a few percent.

\subsection{Kepler Solver}\label{sec:kepler}
Since no closed form solution to Kepler's equation exists, this non-linear problem must be solved iteratively.
In the first version of \WHFastFiveTwelve, an initial guess was refined using a fixed number of iterations.
This works for packed systems such as the Solar System because we can assume a priori that the planets occupy only a restricted range of orbital parameters (eccentricity and orbital period), as they would otherwise collide with one another.
However, this approach is not always optimal and can fail for some systems, particularly those in which high eccentricities might occur temporarily without leading to an instability.
For this reason, we improve the Kepler solver in this version of \WHFastFiveTwelve $\,$ to make it more general.

We follow the notation of \cite{Mikkola1997} and \cite{ReinTamayo2015} and write Kepler's equation as
\begin{eqnarray}
    r_0 X + \eta_0\; G_2(\beta,X) + \zeta_0\; G_3(\beta,X) - \mathit{dt} = 0, \label{eq:kepler} 
\end{eqnarray}
where $G_n$ are Stiefel functions \citep{Stiefel1975}
\begin{eqnarray}
    G_n(\beta, X) = X^n c_n(\beta\, X^2). \label{eq:G} 
\end{eqnarray}
The Stiefel functions in turn depend on Stumpff functions $c_n$ which can be written as a Taylor series (see Eq.~\ref{eq:c}).
The remaining quantities are  
\begin{eqnarray}
    \beta = \frac {2M}{r_0} - v_0^2,\quad\;     \eta_0 = \mathbf{r}_0\cdot\mathbf{v}_0, \;\;\;\mathrm{and}\;\;\;  \zeta_0 = M - \beta r_0, \nonumber
\end{eqnarray}
which depend on the initial positions $\mathbf{r}_0$, velocities $\mathbf{v}_0$, and the standard gravitational parameter\footnote{We write the standard gravitational parameter as $M$ in this paper because without loss of generality we can set the gravitational constant to one and thus reduce the number of multiplications. See also Section~\ref{sec:units}.}, $M$.
The goal of the Kepler solver is to find $X$ given the timestep~$dt$, $M$, and the positions and velocities at the beginning of the timestep.

Our new algorithm works as follows.
We start with a first order guess for $X=dt/r_0$.
We then perform two Halley steps to refine $X$.
In the first step, we include only 9 terms in the Taylor expansion of the Stumpff functions.
In the second step, we include 11 terms.
Now that the estimate is relatively close to the solution, we refine it with a Newton step for which we include 19 terms in the Stumpff functions.
At this point we introduce a convergence check, similar to what is present in \WHFast but is absent in the first version of \WHFastFiveTwelve.
We refine the solution for each planet with Newton iterations until it has converged according to an empirically determined criterion $|\delta X| <\epsilon=10^{-11}$.
We found that if we have not yet converged after four Newton iterations, then we are likely to not converge at all.
Thus, if a converged solution is not obtained after the first four Newton iterations, then we switch to a fallback bisection method.
Once the solution for $X$ has converged, we calculate the $f$ and $g$ functions:
\begin{eqnarray}
    f = M \frac{G_2}{r_0}\quad\quad && \dot f = \frac{M\,G_1}{r_0r} \\
    g = dt - M G_3       \quad\quad     && \dot g = \frac{M\,G_2}{r},
\end{eqnarray}
where $r$ is the distance at the end of the timestep, $r = r_0 + \eta_0 G_1 + \zeta_0 G_2$.
Note that our definitions of $f$, $\dot f$, and $\dot g$ (but not $g$) differ from the definitions in \cite{Mikkola1997}.
Using our definitions, the positions and velocities at the end of the timestep can be written as
\begin{eqnarray}
    \mathbf{r} = (1-f) \mathbf{r}_0 + g \mathbf{v}_0, \quad\;\;
    \mathbf{v} = - \dot f \mathbf{r}_0 + (1 - \dot g) \mathbf{v}_0. 
\end{eqnarray}
Our definitions of $f$, $g$, $\dot f$, and $\dot g$ allow us to use more efficient and accurate fused multiply-add instructions in this final step and write it as
\begin{eqnarray}
    \mathbf{r} &=& \mathrm{fma}(g, \mathbf{v}_0, \mathrm{fnma}(f,\mathbf{r}_0, \mathbf{r}_0)) \\
    \mathbf{v} &=& \mathrm{fnma}(\dot f, \mathbf{r}_0, \mathrm{fnma}(\dot g,\mathbf{v}_0, \mathbf{v}_0))
\end{eqnarray}
where $\mathrm{fma}(a,b,c)=ab+c$ and $\mathrm{fnma}(a,b,c)=-ab+c$.

With AVX512, we can perform the Kepler step in parallel for eight planets.
However, different planets might require a different number of iterations to converge, with typically the innermost or most eccentric planet requiring the most.
To ensure that each planet's solution remains independent of all other planets in the simulation, only the lanes for planets that have not yet converged are updated at each iteration and planets that have already converged are no longer modified.
This is important because an excessive number of iterations on an already converged solution can actually worsen the accuracy due to accumulating roundoff errors.
We achieve this using masked instructions that write only to the lanes corresponding to planets that have not yet converged.
Masked instructions themselves have no overhead in terms of latency and throughput, but we require some CPU cycles to generate and update the masks at each iteration.

The specific sequence of Halley and Newton steps and the number of terms used in the Taylor series expansion of the Stumpff functions were identified through a brute-force comparison of all possible combinations, subject to the requirement of convergence to machine precision.
We then chose the most efficient sequence for what we consider a typical integration (see Appendix~\ref{app:bruteforce} for more details).
For different applications, for example in cases with highly eccentric orbits, a different sequence might be slightly faster.

The Newton steps do not converge quickly for highly eccentric or hyperbolic orbits.
In that case we use a fallback bisection method that is identical to that of \WHFast \citep{ReinTamayo2015}.
The bisection method is guaranteed to converge but is significantly slower than the two or three Newton steps we require for low and moderately eccentric orbits.
However, we do not expect highly eccentric and in particular hyperbolic orbits to persist for very long in a typical simulation.

To summarize our implementation of the Kepler solver, we list the pseudocode in Algorithm~\ref{alg:kepler}.
The Kepler solver is evaluated for all 8~planets in parallel using AVX512 instructions.
Note that for readability, we do not show the masked instructions or the mask generation.

\begin{algorithm}[t]
    \setlength{\algomargin}{0pt}
	\hrule
	\vspace{4pt}
    \DontPrintSemicolon
	\KwIn{$\mathit{dt}$, positions, velocities}
	\KwOut{positions, velocities}
    $r, \beta, \eta, \zeta \gets $ from positions and velocities\\
    \tcc{First order guess}
    $\assign{X} \mathit{dt}/r$ \\
    \tcc{First two refinements}
    $\assign{G_{0,1,2,3}}$ {Stiefel}($X, \beta$, nTerms = 9)\\
    $\assign{X} $ {Halley}($X, \mathit{dt}, G_{0,1,2,3}, \beta, \eta$)\\
    $\assign{G_{0,1,2,3}} $ {Stiefel}($X, \beta$, nTerms = 11)\\
    $\assign{X} $ {Halley}($X, \mathit{dt}, G_{0,1,2,3}, \beta, \eta$)\\
    \tcc{Newton iteration loop}
    $\assign{X'} $ $X$\\
    \For{$i\leftarrow 0$ \KwTo $4$}{
        $\assign{\mathit{G_{1,2,3}}} $ {StiefelCompensated}($X, \beta$, nTerms = 19)\\
        $\assign{X} $ {Newton}($X, \mathit{dt}, G_{1,2,3}, \beta, \eta$)\\
        \If{$|X-X'|<\epsilon$}{
            \textbf{goto} converged
        }
        $\assign{X'} $ $X$\\
    }
    \tcc{Fallback bisection method}
    \If{$\beta \geq 0$}{
        $\text{Period} \gets $ from positions and velocities\\
        $\assign{X_{min}}  2\pi \cdot \;\text{floor}(\mathit{dt}/\text{Period}) / \sqrt{\beta}$ \\
        $\assign{X_{max}}  X_{min} + 2\pi / \sqrt{\beta}$ \\
        $\assign{X}  (X_{min} + X_{max})/2$ \\
        \For{$i\leftarrow 0$ \KwTo $51$}{
            $\assign{\mathit{G_{1,2,3}}} $ {Stiefel}($X, \beta$, nTerms = 19)\\
            \eIf{$r\cdot X + \eta \cdot G_2 + \zeta \cdot G_3 < \mathit{dt}$}{
                $X_{min} \gets  X$ \\
            }{
                $X_{max} \gets X$ \\
            }
            $\assign{X}  (X_{min} + X_{max})/2$ \\
        }
    }
    \Else{
        \tcp{Hyperbolic orbit}
        $v_q, q \gets $ from positions and velocities\\
        $\assign{X_{min}} \mathit{dt}/(|v_q\cdot \mathit{dt}| + r_0)$ \\
        $\assign{X_{max}} \mathit{dt}/q$ \\
        \While{$|X_{max}-X_{min}|>10^{-15}$}{
            $\assign{\mathit{G_{1,2,3}}} $ {Stiefel}($X, \beta$, nTerms = 19)\\
            \eIf{$r\cdot X + \eta \cdot G_2 + \zeta \cdot G_3 < \mathit{dt}$}{
                $X_{min} \gets  X$ \\
            }{
                $X_{max} \gets X$ \\
            }
            $\assign{X}  (X_{min} + X_{max})/2$ \\
        }
    }

    \tcc{Final update}
    converged:\\
    $\assign{G_{1,2,3}} $ {StiefelCompensated}($X, \beta$, nTerms = 19)\\
    ${f, g, \dot f, \dot g} \gets $ using $G_{1,2,3}, \eta, \zeta, \mathit{dt}$\\
    new positions $\gets$ using ${f, g}$, positions, velocities \\
    velocities $\gets$ using ${\dot f, \dot g}$, positions, velocities \\
    positions $\gets$ new positions \\
	\vspace{4pt}
	\hrule
    \begin{minipage}{\linewidth}
    \caption{Pseudocode of the new \WHFastFiveTwelve Kepler solver. 
        $G_{0,1,2,3}$ are the Stiefel functions defined in \cite{Mikkola1997} and implemented using Algorithm~\ref{alg:stiefel}. 
        The quantities $X$, $\beta$, $\eta$, and $\zeta$ are defined in \cite{ReinTamayo2015}.~$q$~and $v_q$ are the periapsis distance and speed, respectively.
        \label{alg:kepler}}
    \end{minipage}
\end{algorithm}

\subsection{Roundoff in the Stiefel functions} \label{sec:roundoff}
We need to evaluate Stiefel functions $G_n$ for the Kepler solver described above.
The Stiefel functions in turn depend on the Stumpff functions $c_n$.
We calculate the Stumpff functions from a truncated Taylor series:
\begin{eqnarray}
    c_n(z) \equiv \sum_{j=0}^{n_{\rm terms}} \frac{(-z)^j}{(n+2j)!} \label{eq:c}.
\end{eqnarray}
In floating point arithmetic, this evaluation carries a small but systematic bias.
The bias has two different origins.
First, the inverse factorial coefficients of the series are pre-calculated and rounded to the nearest representable number.
Second, the summation of the series involves rounding at every term.
Both effects are tiny, of the order of $10^{-20}$ per timestep, but they can accumulate over time because the rounding errors in the coefficients are constant.
This causes the energy error in long integrations to grow linearly in time.

This bias is present in the first version of \WHFastFiveTwelve, although it is hardly visible in a typical simulation because the overall scheme error is significantly larger than in our new version (see Section~\ref{sec:jacobi}).
Here, we aim to implement an even less biased scheme in which the energy error grows only with the square root of the number of steps, consistent with Brouwer's law \citep{Brouwer1937}.

We reduce the bias with the following two changes.
First, we precalculate the rounding errors of the inverse factorial coefficients of the series in extended precision.
Second, we evaluate the series with a compensated summation that includes both the original inverse factorial coefficients and rounding error corrections.
The compensated summation retains the low order bits that would otherwise be lost.
Neither change is sufficient on its own because the coefficient corrections are too small to be represented in double precision and are simply discarded unless the compensated summation also carries the extra bits.
We test the effectiveness of both changes together as well as individually in Appendix~\ref{app:roundoff_ablation}.
Only the last few terms of the series require this correction because, for a typical timestep, the correction is small and the higher order terms are negligible.
We only use the correction for the Newton iterations, not the initial two Halley steps.
This correction adds about 5\% to the total runtime but it restores the expected random walk behaviour of the energy error over multi-Gyr integrations as we show in Section~\ref{sec:longterm} even for orbits with very high eccentricities (see in Appendix~\ref{app:highe}).

We list the pseudocode for our Stiefel function evaluation in Algorithm~\ref{alg:stiefel}.
The Stiefel functions are evaluated for all 8~planets in parallel using AVX512 instructions.

\begin{algorithm}[t]
    \setlength{\algomargin}{0pt}
	\hrule
	\vspace{4pt}
    \DontPrintSemicolon
	\KwIn{$X$, $\beta$, nTerms}
	\KwOut{$G_{0,1,2,3}$}
    $\assign{z} \beta \cdot X^2$\\
    \tcc{Stumpff functions by Horner's method}
    $\assign{c_2} 1/(\mathrm{nTerms}-1)!$\\
    $\assign{c_3} 1/\mathrm{nTerms}!$\\
    \tcc{Plain Horner for the higher order terms}
    \For{$k \leftarrow \mathrm{nTerms}-2,\, \mathrm{nTerms}-4,\, \ldots,\, 7$}{
        $\assigntwo{c_3} 1/k! - z\cdot c_3$\\
        $\assigntwo{c_2} 1/(k-1)! - z\cdot c_2$\\
    }
    \tcc{Compensated Horner for last two terms}
    $\assign{d_2} 0$\\
    $\assign{d_3} 0$\\
    \For{$k \leftarrow 5,\, 3$}{
        \vspace{1mm}
        $\assigntwo{p} z\cdot c_3$\\
        $\assigntwo{p_e} \mathrm{fma}(z, c_3, -p)$\\
        $\assigntwo{c_3} 1/k! - p$\\
        $\assigntwo{s_e} (1/k! - c_3) - p$\\
        $\assigntwo{d_3} (s_e - \delta_k - p_e) - z\cdot d_3$\\
        $\assigntwo{p} z\cdot c_2$\\
        $\assigntwo{p_e} \mathrm{fma}(z, c_2, -p)$\\
        $\assigntwo{c_2} 1/(k-1)! - p$\\
        $\assigntwo{s_e} (1/(k-1)! - c_2) - p$\\
        $\assigntwo{d_2} (s_e - \delta_{k-1} - p_e) - z\cdot d_2$\\
    }
    $\assign{c_2} c_2 + d_2$\\
    $\assign{c_3} c_3 + d_3$\\
    \tcc{Stiefel functions from Stumpff functions}
    $\assign{G_2} X^2\cdot  c_2$\\
    $\assign{G_3} X^3\cdot  c_3$\\
    $\assign{G_1} X - \beta\cdot  G_3$\\
    $\assign{G_0} 1 - \beta\cdot  G_2$\\
	\vspace{4pt}
	\hrule
    \begin{minipage}{\linewidth}
        \caption{Pseudocode of our Stiefel function evaluation with compensated summation. The Stumpff functions $c_2$ and $c_3$ are evaluated by Horner's method from their truncated Taylor series, and the Stiefel functions follow from the recurrence relations~\citep{Mikkola1997}. For the last two terms that appear in Horner's method, compensated summation is used to accumulate the rounding of every operation in a low order word $d$, which is added back at the end. Here, $\mathrm{fma}(a,b,c)=ab+c$ is evaluated with a single rounding, and $\delta_k$ is the precomputed rounding error of the coefficient $1/k!$. This compensation is applied only in the final evaluation that produces the~$f$ and~$g$ functions; intermediate evaluations use plain Horner for all terms.
\label{alg:stiefel}}
    \end{minipage}
\end{algorithm}

\section{Optimizations}\label{sec:optimizations}
In Section~\ref{sec:algo} we described algorithmic improvements to \WHFastFiveTwelve.
In this section we focus on the implementation details that make this algorithm run fast.

\subsection{Assembly}
The first version of \WHFastFiveTwelve was written in C99 with special intrinsics that map directly to AVX512 instructions.
This approach makes it easy to include AVX512 in existing code.
However, one disadvantage is that it does not allow fine-grained control over various aspects of the code because intrinsics leave those details to the compiler.
For example, the compiler allocates registers and decides which variables to put on the stack.

For these reasons, the main parts of the new version of \WHFastFiveTwelve are now written in about 800 lines of x86~assembly.
We make extensive use of macros, which allow us to reuse much of the code for different configurations.
Specifically, macros generate 24 slightly different functions for simulations with different number of planets, with and without general relativistic corrections, with and without encounter checks, as well as with and without ejection checks (see Section~\ref{sec:gr}).
Similarly, macros create six different functions for symplectic correctors.
This approach allows us to provide highly optimized functions for a variety of use cases while avoiding branching because we can select the most appropriate kernel once at the beginning of the integration.

\subsection{Keeping the Simulation in Registers}
AVX512 supports 32 CPU registers, each 512 bits wide.
Using only seven registers, we can keep the entire simulation state, specifically the masses, positions, and velocities of all particles, in registers at all times.
The remaining registers are sufficient for performing all computations and storing intermediate values and constants such as the timestep and numerical factors.

We still need to use memory to store some constants, such as the 128~matrix coefficients (see below) as these alone would take up half the available registers.
However, by keeping the simulation state in registers, we never have to write to memory during an integration except when an output is required.
Thus, the cache is not polluted and can work very efficiently.
In fact, we have zero cache misses\footnote{Cache misses might still occur if the operating system performs some other operation on the CPU during an integration.}.
As a result, \WHFastFiveTwelve's performance is not limited by the speed of memory access in contrast to many other HPC applications.

\subsection{Matrix Multiplications}\label{sec:matrix}
As discussed in Section~\ref{sec:jacobi}, we need to perform several matrix multiplications to transform from Jacobi coordinates to heliocentric coordinates and back.
In the case of eight planets, we need to multiply an~$8\times8$~matrix by an $8$-element vector containing $x$, $y$, or $z$ coordinates.
Thus, in total, we need to perform six matrix-vector multiplications per timestep.
It is therefore crucial to optimize this calculation as much as possible.
We achieve very good performance by combining several approaches.

First, the matrix elements are constants stored in memory.
Because we never write to memory (see above), they remain in the CPU's L1~cache, and load operations are therefore very efficient.

Second, we perform the multiplications for three coordinates at the same time.
This way, we need to load each matrix element only once and can reuse it for all three coordinates.

Third, we use two accumulators per coordinate. The accumulators are only combined at the end of the multiplication.
This allows the CPU to schedule instructions in parallel and avoids data dependencies in the pipeline.

Fourth, we use fused multiply-add instructions extensively. These instructions improve both throughput and accuracy by avoiding an intermediate rounding step.

\subsection{Avoiding Branches}
Branches such as those originating from conditional statements or loops can significantly slow down a simulation.
This is because modern CPUs often execute instructions out-of-order, but branches with multiple possible outcomes make such execution more difficult.
We therefore carefully remove almost all branches in \WHFastFiveTwelve's kernel and are left with only two.

The first branch controls the overall loop over individual timesteps.
This branch allows the loop to terminate after the specified number of timesteps.
The other remaining branch is in the Kepler step, where it checks for convergence (see Section~\ref{sec:kepler}).

\begin{figure*}[t]
    \includegraphics[trim=0 8mm 0 0, clip, width=\textwidth]{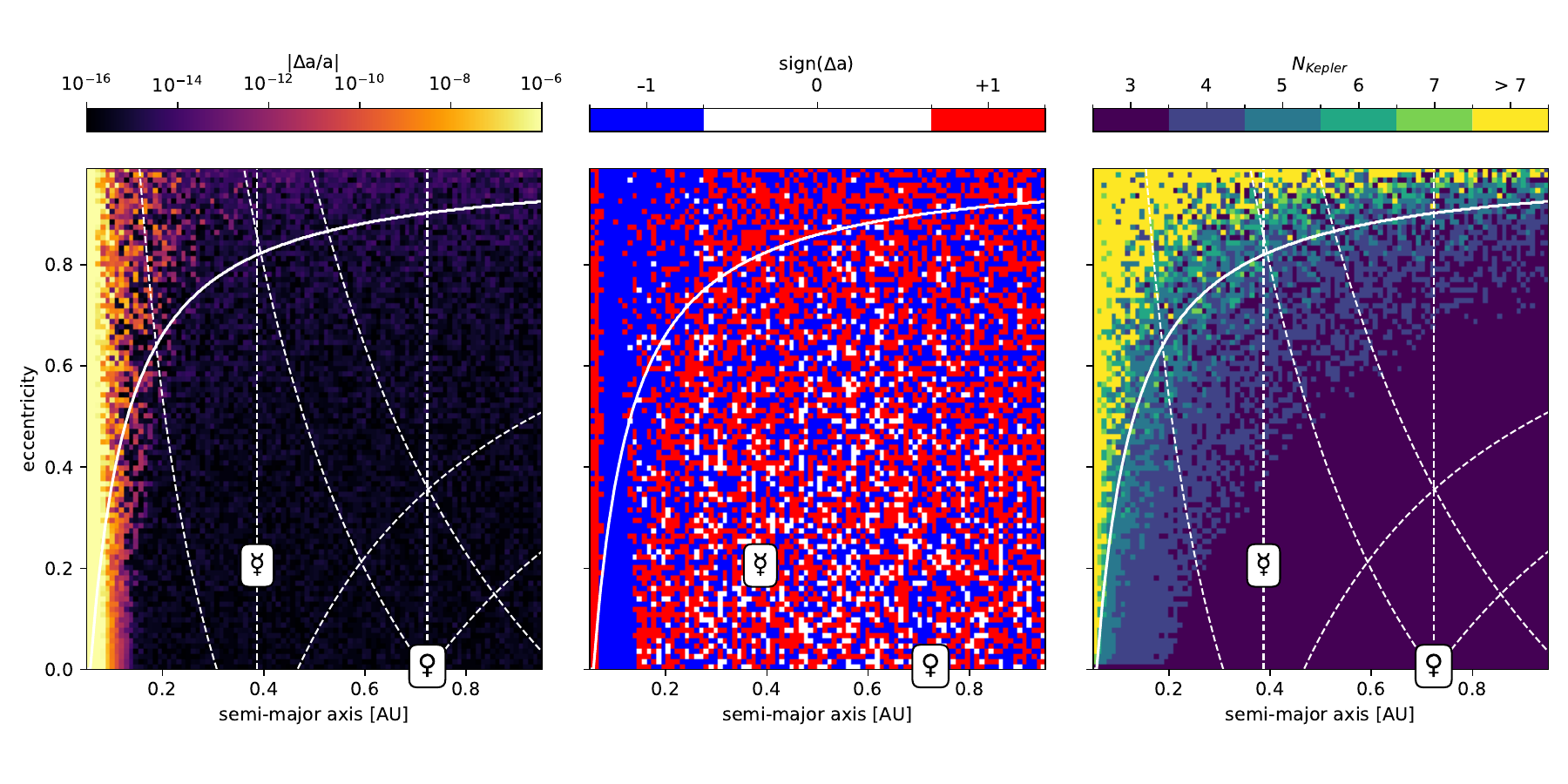} 
    \caption{
        The absolute relative semi-major axis error, the sign of the semi-major axis error, and the number of iterations in the Kepler solver.
        Here, a test particle is integrated for one timestep of $5$~days. The current locations of the Solar System planets are shown in each plot. See the text for a description of the lines.
        \label{fig:kepler}}
\end{figure*}

\subsection{Square Roots}
The interaction step is dominated by the pairwise force evaluations between planets, each of which is proportional to $1/r^3$, where $r$ is the distance between the two particles.
No single instruction computes $1/r^3$, and on current CPUs and both the square root and the division instruction have a low throughput and high latency.
We therefore use the approximate reciprocal square root instruction \texttt{rsqrt14}, which returns an estimate of $1/\sqrt{x}$ accurate to about 14~bits.
The \texttt{rsqrt14} instruction has a latency of about 8 clock cycles compared to over 50 for the normal square root plus division instructions.
We already have $r^2$ from the distance calculation, so applying \texttt{rsqrt14} to $r^2$ returns an estimate of $1/\sqrt{r^2}=1/r$.
Similar to \cite{Tanikawa2012}, we set $a=r^2$ and use Newton-Raphson iterations to solve $f(y)=y^{-2}-a=0$, whose positive root is $y=1/\sqrt{a}=1/r$. Starting from the \texttt{rsqrt14} estimate $y_0$, each iteration is $y_{n+1}=y_n(3/2-a y_n^2/2)$. We then obtain $1/r^3=y_2^3$ with two additional multiplications. Doing so makes the reciprocal square root accurate to machine precision.
This procedure replaces the expensive square root and the division instructions with a short sequence of multiply and fused-multiply-add instructions, all of which are high throughput and low latency.

In comparison to the actual IEEE~754 compliant square root and division instructions, the result of our procedure is not guaranteed to be correctly rounded.
However, in the case of the interaction step, the rounding is not crucial as a) the accelerations we calculate are small to begin with as they are perturbations to the Keplerian motion, b) because they get multiplied by a small number, the timestep, and c) because we encounter the same rounding error for both particles and thus, thanks to Newton's third law, energy is conserved regardless.

To illustrate the last point, let us consider the interaction between two planets at a distance $r$ from each other.
Suppose we calculate everything exactly but incorrectly round the inverse square root to $\left[1/\sqrt{r}\right]_{\rm fp} = 1/\sqrt{r} + \delta$ where $\left[\cdot\right]_{\rm fp}$ represents our approximation and $\delta$ is the error.
Then, with the exception of the sign, the error in the force felt by both planets is exactly the same as long as we use the same approximation $\left[1/\sqrt{r}\right]_{\rm fp}$ for both planets.
A different way to interpret this is that we actually calculate forces without \textit{any} error (and most importantly ensuring $F_{ij}=F_{ji}$), just for a slightly wrong force law.

In the Kepler step we need both the distance $r$ as well as its reciprocal $1/r$: $r$ is used to compute $\zeta$, and $1/r$ to compute $\beta$ and the first order guess for $X$.
Here the square root is evaluated only once per planet and timestep, and not once for each of the $N\!\cdot\!(N\!-\!1)/2$ pairs as in the interaction step.
It therefore does not lie on the throughput-critical path, and the Kepler step remains dominated by the later iterative solver.
Furthermore, because the result of the square root operation is used at the end of the Kepler step to advance the positions and velocities, any bias would propagate to those quantities.
We thus avoid the \texttt{rsqrt14} instruction used for the pairwise forces and use the IEEE~754 compliant square root and division instructions in the Kepler step instead.

\section{Tests}\label{sec:tests}

\subsection{Kepler solver}\label{sec:testkepler}
\begin{figure*}[t]
    \centering
\includegraphics[trim=0 2mm 0 0, clip, width=0.68\textwidth]{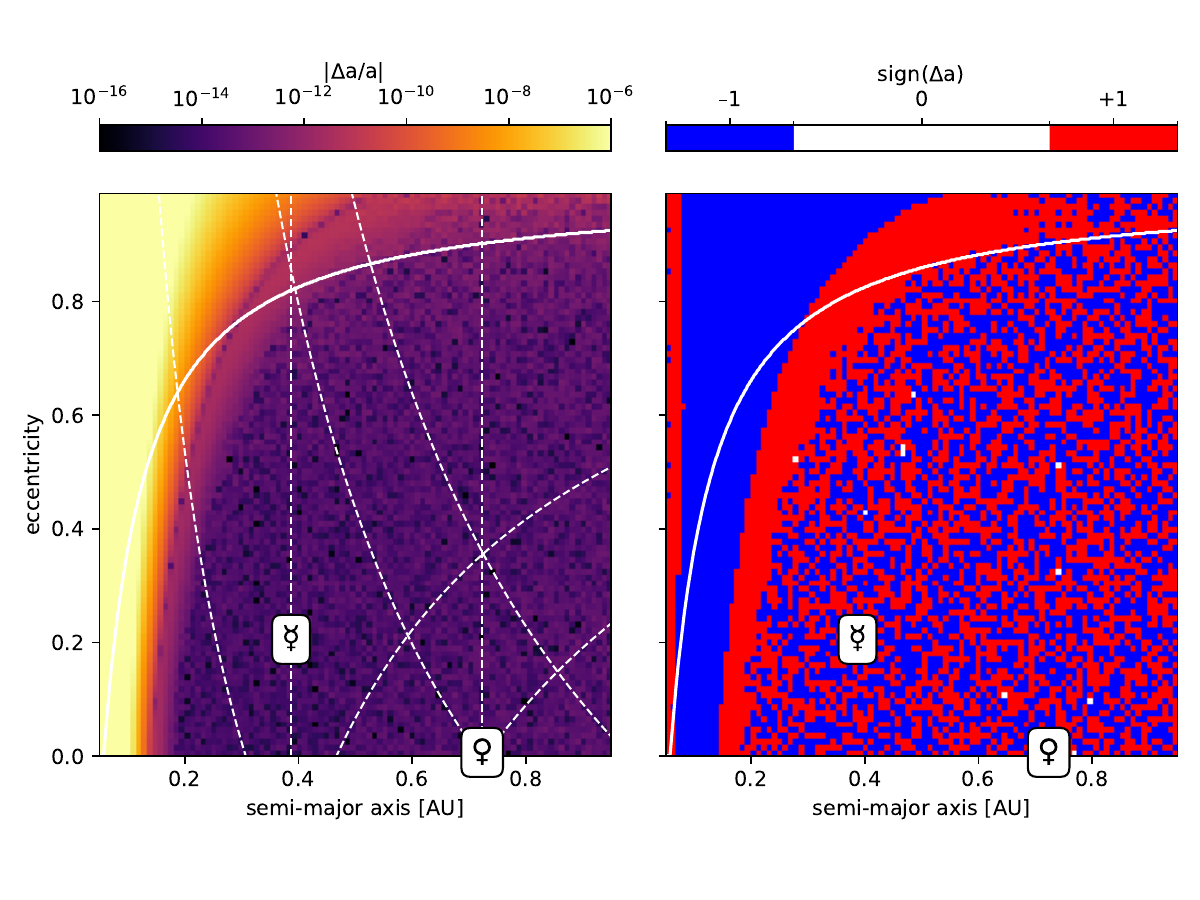} 
    \vspace{-0.6cm}
\caption{
    Same as the first two panels of Fig.~\ref{fig:kepler}, but each particle is integrated for 1000~orbital periods. The top and left of the plots show a biased error growth. The bottom right of the plots show an unbiased error growth in the most important area of parameter space.
    \label{fig:kepler_long}}
\end{figure*}

The Kepler solver of \WHFastFiveTwelve is crucial for both the speed and accuracy of the entire integrator.
We therefore test the Kepler solver extensively with one particle at a time.
We use a fixed timestep of $5$~days and plot the results over a grid of semi-major axis and eccentricity.
The phase of the orbit is chosen randomly.
Because gravity is scale invariant, our choice of a timestep of $5$~days is somewhat arbitrary and we chose it simply because similar values have been used in past simulations of the Solar System.
What actually matters is the ratio of the timestep over the orbital period.
Because of that the results of this test can be applied to any other planetary system.

Figure~\ref{fig:kepler} shows the relative semi-major axis error, the sign of the semi-major axis error, as well as the number of iterations in the Kepler solver.
We also overplot the current orbital parameters of Mercury and Venus. The vertical dashed lines correspond to the planets' semi-major axes.
The curved dashed lines can be used to find regions in parameter space in which a particle crosses the orbit of a Solar System planet on its present-day orbit.
For example, if the eccentricity of Mercury rises to $\sim0.85$, then its orbit will cross the present-day orbit of Venus.
Similarly, if the eccentricity of Venus rises to 0.4, then it will cross orbits with both Mercury and Earth.
The solid white line shows where the timestep is equal to the pericenter timescale $T_f = 2\pi(1-e)^2/\left(n\sqrt{1-e^2}\right)$, where $e$ is the eccentricity and $n$ the mean motion.
These lines help to understand whether the accuracy in a region of parameter space matters or not.
For example, if the timestep is larger than the pericenter timescale (top left above the solid white line), then we most definitely fail to resolve physically relevant timescales\footnote{\cite{Wisdom2015} finds that at least 17~timesteps per pericenter timescale are required for some systems.}.
Similarly, if a planet becomes very eccentric, then the planet is likely to have a close encounter with a neighbouring planet shortly afterwards.
In both cases, the Kepler solver is no longer the limiting factor for the accuracy of a simulation.

The left panel of Fig.~\ref{fig:kepler} shows that, in the most relevant region of parameter space, the Kepler solver is accurate to machine precision, $\sim\!10^{-16}$.
The middle panel shows that the sign of the error is random in the same region.
This is important for having an unbiased integrator (see Section~\ref{sec:kepler} and Appendix~\ref{app:roundoff_ablation}).
For example, imagine we have an integrator accurate to $10^{-16}$ after one timestep but with a bias of the same magnitude (i.e., the sign of the error is the same every time).
After, say, $10^{11}$ steps corresponding to a few Gyr in a Solar System integration, the error in the semi-major axis would have increased to $10^{-5}$ from this error source alone.
If, on the other hand, the integrator is unbiased, then the error would only be~$10^{-16}\cdot 10^{11/2} \approx 3\cdot10^{-11}$.

The right panel shows the number of iterations required for the Kepler solver to converge.
Since we always have two Halley steps followed by at least one Newton step, the minimum number of iterations is~3.
For a large region of parameter space, we indeed converge in only~3~iterations.
More iterations are required for higher eccentricities and shorter orbital periods (we provide a direct comparison with \WHFast for bound orbits with extremely high eccentricities in Appendix~\ref{app:highe}).
The yellow colour indicates where we fail to converge in 7~iterations and thus have to fall back to the bisection method.
This only occurs for very eccentric orbits. 
Note that even in the case where the bisection method was used, the error remains close to machine precision and the sign of the error remains random.

The Kepler solver fails for very short orbital periods.
This is expected because our algorithm does not support timesteps that are larger than the orbital period.
Fixing this is easy enough to do (it is done in \WHFast), but would introduce an additional branch in the code which would affect the speed of the algorithm.
Since simulations in which the timesteps are large compared to the innermost planet's period lead to unphysical results anyway, we decided to not implement a Kepler solver which unconditionally converges in all cases.

We next repeat the tests from Fig.~\ref{fig:kepler} with the same 5~day timestep but now integrate each particle for 1000~orbital periods instead of only one timestep.
The results are shown in Fig.~\ref{fig:kepler_long}.
As expected, the relative semi-major axis error is now larger as errors accumulate.
The sign of the semi-major axis error is still random in the bottom right region of parameter space, indicating that the Kepler solver remains unbiased here.
However, the area in which the sign is consistently positive or negative has increased.
This result reveals a small but biased contribution to the error.
As this term grows $\propto t$, it will dominate over the unbiased $\propto \sqrt t$ term over sufficiently long timescales.
We show in Section~\ref{sec:longterm} that the timescale over which this bias becomes apparent in a simulation of the Solar System is much longer than the age of the Solar System.

\subsection{Convergence}\label{sec:convergence}
\begin{figure}[t]
    \includegraphics[trim=0 2mm 0 0, clip, width=\columnwidth]{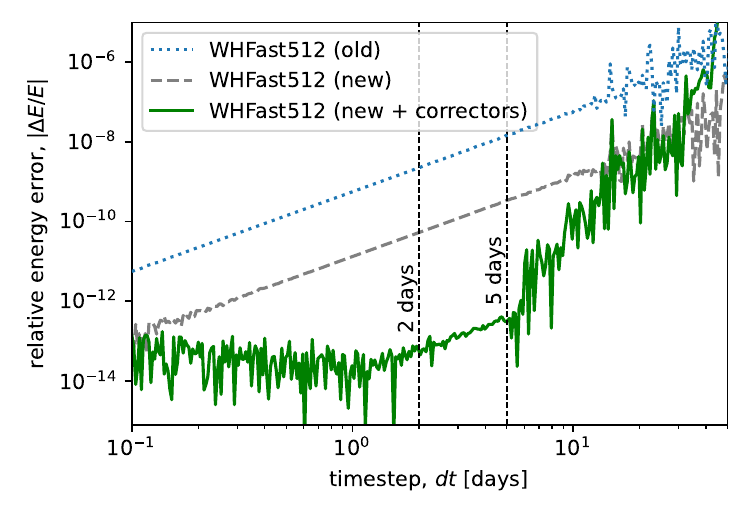} 
    \caption{
        Relative energy error as a function of the timestep for different integrators in a 100~year simulation of the Solar System with all eight planets.
        \label{fig:convergence}}
\end{figure}

We test the convergence of \WHFastFiveTwelve by integrating the Solar System with all eight planets for 100~years.
The relative energy error at the end of the integration is plotted as a function of the timestep in Fig.~\ref{fig:convergence} for different integrators.

The new \WHFastFiveTwelve integrator converges as the timestep gets smaller, as expected for a second order scheme.
When symplectic correctors are turned on, the errors are reduced by a further factor of $\sim\! 10^{3}$ for timesteps~$\sim\!5$~days.
For timesteps smaller than $\sim\!2$~days, the error starts to increase again as the timestep is reduced further because we reach the floor imposed by double precision floating point arithmetic.

Note that the old version of \WHFastFiveTwelve performs significantly worse because it uses democratic heliocentric coordinates rather than Jacobi coordinates and does not support symplectic correctors.
For the same timestep, the new version of \WHFastFiveTwelve is~5~orders of magnitude more accurate.

\subsection{Long-term Integrations}\label{sec:longterm}
\begin{figure}[t]
    \includegraphics[trim=0 2mm 0 0, clip, width=\columnwidth]{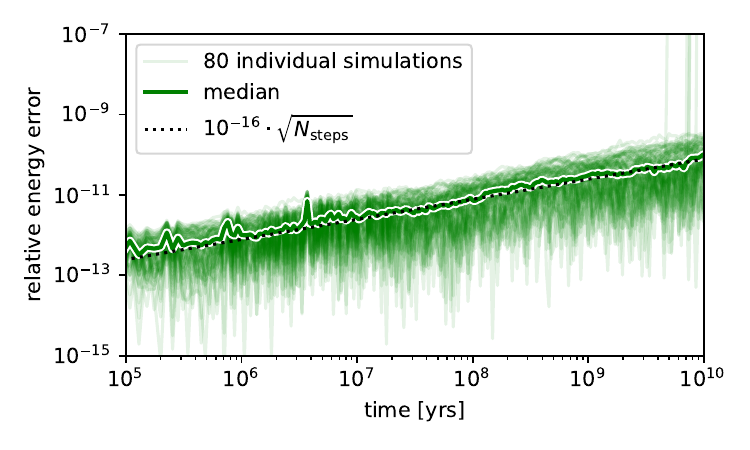} 
    \caption{
        The relative energy error as a function of time in 80 simulations of the Solar System. The bold green line shows the median. The dotted line shows the expected behaviour for an unbiased integrator that is accurate to machine precision after one timestep.
        Several simulations go unstable at late times, leading to a large energy error.
        \label{fig:longterm}}
\end{figure}

We test the long-term conservation properties of \WHFastFiveTwelve by simulating the Solar System with all eight planets and general relativistic corrections for~10~Gyr.
We use a timestep of 6~days and turn on symplectic correctors.
The initial conditions of each simulation are perturbed by a few meters in a random direction.
The relative energy errors of the 80 simulations and their median are shown as a function of time in Fig.~\ref{fig:longterm}.

The median relative energy error remains below $10^{-10}$ over 10~Gyr.
The errors grow as $\Delta E/E \sim 10^{-16}\cdot \sqrt{N_{\rm steps}}$.
This shows that the integrator is unbiased and follows Brouwer's law \citep{Brouwer1937} over physically relevant timescales.
We expect that over much longer timescales, we will eventually see linear error growth (see the discussion in Section~\ref{sec:testkepler}).

Note that several simulations go unstable at late times which leads to a spike in their energy errors.
Instabilities in the Solar System are expected and the rate of instability is consistent with previous results \citep{LaskarGastineau2009, Rein2026}.

\subsection{Runtime}

To measure the runtime, we compare \WHFastFiveTwelve to \WHFast, one of the fastest publicly available N-body integrators for planetary systems \citep{Javaheri2023}.
Note that \WHFast has not been manually optimized for AVX512 but a compiler could in principle try to optimize it with AVX512 instructions automatically. 
We once again use the Solar System with all 8 planets, a $6$~day timestep, and general relativistic corrections as the canonical test case. 
All results presented here are from simulations performed on an AMD EPYC 9655 2.6GHz processor. 
Similar speedups were measured on an Intel Xeon Gold 6148 2.4GHz CPU.

We were able to finish a 5~Gyr integration in only 9.0~hours. 
This corresponds to a 60\% speedup when compared to a simulation with the earlier version of \WHFastFiveTwelve using the same timestep and CPU.
When compared to the non-SIMD version of \WHFast, this corresponds to a more than $8\times$ speedup.
For \WHFast, aggressive compiler optimizations were used (\texttt{-O3}, \texttt{-march-native}).
For \WHFastFiveTwelve the specific compiler or compiler optimizations do not matter as it is written in assembly.

\begin{figure}[t]
    \includegraphics[width=\columnwidth]{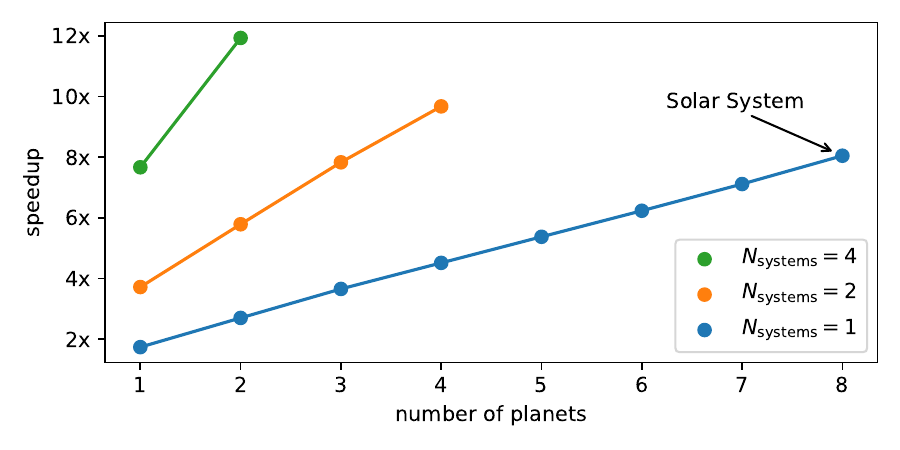} 
    \caption{
        The speedup of \WHFastFiveTwelve compared to \WHFast in simulations with different number of planets. 
        For \WHFastFiveTwelve simulations that use $N_{\rm systems}>1$, the corresponding \WHFast simulation are run sequentially.
        \label{fig:Nsystems}}
\end{figure}

If the number of planets $N$ is $\leq4$ \WHFastFiveTwelve can integrate multiple systems at the same time as we can pack one AVX512 register with data from up to 8~planets and they don't necessarily need to be in the same system.
The speedup for different numbers of planets and different numbers of planetary systems integrated at the same time are shown in Fig.~\ref{fig:Nsystems}.
As one can see, the speedup is linear in the number of planets if the number of systems is kept fixed. 
\WHFastFiveTwelve performs best when 8 planets are integrated at the same time as this fills up one AVX512 register.
For the case of 2 planets in 4 different planetary systems, \WHFastFiveTwelve is more than $12$~times faster than \WHFast.

\subsection{A Note on Units}\label{sec:units}
In the tests in this section, we use physical units to describe the setup and results.
For example, we use a timestep of 5~days.
Since gravity is scale invariant, this does not lead to a loss of generality.
What matters physically is the timestep relative to the orbital period.

However, we introduce a scale into the problem by setting the Kepler solver's convergence criterion to a fixed value $\epsilon$, which has units of time divided by length.
In practice, this means that the units should be chosen such that positions and velocities are not excessively large or small.
For example, having the innermost planet at 0.01~AU or 100~AU is perfectly fine, but having a planet at $10^{11}$m or $10^{-12}$Mpc is not.
If necessary, $\epsilon$ can to be adjusted accordingly.

Furthermore, \WHFastFiveTwelve is hardcoded to use units in which the gravitational constant is equal to one.
This allows us to avoid multiplications with the gravitational constant entirely.
We also hardcode the speed of light when general relativistic corrections are used.
Thus, in practice and without loss of generality, the code units used by \WHFastFiveTwelve measure length in astronomical units and time in~${\rm yr}/2\pi$.

\section{Discussion}\label{sec:discussion}
We argue that, with the improvements described in Sections~\ref{sec:algo} and \ref{sec:optimizations}, we approach the optimal performance attainable by a Wisdom-Holman type integrator on a present-day high-end CPU.
There are several reasons why further significant improvements (speedups of $\gtrsim 2\times$) will be hard to achieve.

First, the code is entirely compute bound.
Thus, the latency and throughput of instructions, particularly floating point operations, are the limiting factors for \WHFastFiveTwelve.
Since we do not use memory to store the simulation state and read only a small amount of memory for constants and coordinate transformations, the code is never memory bound.
This is in contrast to many other scientific HPC applications where memory access is the main bottleneck.
As a result, higher memory bandwidth, unified memory, in-memory processing, and other memory related technologies are irrelevant to our application.

Second, if we optimize one part of the algorithm, we are limited by the fraction of the total runtime contributed by that part \citep[Amdahl's argument,][]{Amdahl1967}.
In our case, the main parts of the algorithm are the interaction step, the Kepler step, and the matrix-vector multiplications used for coordinate transformations.
These take up approximately 25\%, 63\%, and 11\% of the total runtime, respectively.

Thus, the Kepler step is the most promising target for further optimization.
Within the Kepler step, about 50\% of the time is spent in the iterations solving Kepler's equation.
As we have shown in Section~\ref{sec:testkepler}, we require only three iterations to converge to a solution for most orbits.
We must perform at least two iterations to check for convergence.
Thus, we can remove at most one of the three iterations, for example by finding a better initial guess, using data from previous timesteps, or using a higher order solver.
Even under the optimistic assumption that such a change would not involve any additional computational work, this would give us a speedup of at most $2/3 \cdot 50\% \cdot 63\% \approx 20\%$.
Some improvements might be possible with a different formulation of the Kepler problem itself, for example one that does not rely on the Stiefel functions or the $f$ and $g$ functions.

Next, we consider the interaction step.
Here, the main bottlenecks are the throughput and latency of the square root and division instructions needed for the force calculations.
We already use the faster 14~bit accurate \texttt{rsqrt14} instruction with only two refinement steps.
Thus, without hardware changes, such as providing more floating point units for each CPU core, we do not expect a significant speedup of the interaction step.

Finally, even a substantial improvement to the matrix-vector multiplication would yield an overall speedup of at most~11\%.

For all these reasons, we believe that our implementation of \WHFastFiveTwelve is almost optimal for simulations of planetary systems with $N\leq8$ planets on the current generation of high-end CPUs.

\section{Conclusions} \label{sec:conclusions}
In this paper, we have presented a new version of the \WHFastFiveTwelve N-body integrator.
This Wisdom-Holman style algorithm has been highly optimized for planetary systems with $N\leq8$ planets.
It is by far the fastest $N$-body integrator currently available for this kind of simulation.
Using a timestep of $dt=6$~days, \WHFastFiveTwelve can integrate the eight planets of the Solar System for 5~Gyr in 9.0~hours on an AMD EPYC 9655 processor.
On the same hardware, this is 60\% faster than the previous version of \WHFastFiveTwelve and more than $8\times$ faster than the non-SIMD version of \WHFast, which is itself already significantly faster than most $N$-body codes \citep{Javaheri2023}.

With an almost order of magnitude speed-up, one can explore planetary systems over much longer timescales than what was possible before.
For example, we are now in a position where it is feasible to repeat the secular integrations of \cite{HoangMogaveroLaskar2022} which span 100~Gyr to investigate the dynamical half-life of Mercury with a direct N-body integration that takes just one week.

In an alternative scenario, we can take advantage of the speed-up by running an order of magnitude more simulations than before for the same computational cost.
This allows for much finer parameter space surveys and could for example significantly improve training datasets for machine learning models~\citep{Spock2020}.

We achieved the reported speedup by switching from C to assembly, giving us much finer control over individual registers and CPU instructions.
Specifically, this rewrite allowed us to keep the entire simulation state in CPU registers, reduce the number of branches in the code to just two per timestep, and implement a highly optimized matrix-vector multiplication for the coordinate transformations.

In addition to a significant speedup, we also switched \WHFastFiveTwelve to Jacobi coordinates, allowing us to use symplectic correctors and GR corrections together and thus suppress errors by up to 5~orders of magnitude.
We took great care to make our implementation of the Kepler solver as bias-free as possible, thereby enabling reliable long-term integrations.

Although our discussion has focused on the Solar System as the canonical test case, \WHFastFiveTwelve is a reliable and fast $N$-body integrator for many small-$N$ simulations.
It is particularly well suited for simulations that determine the stability of planetary systems because these simulations are typically stopped as soon as orbits cross or close encounters occur.

We argue that \WHFastFiveTwelve is almost optimal in terms of performance and accuracy for these kind of simulations on present-day CPUs.
Small improvements might still be possible, but the two main bottlenecks are now the square root calculations in the interaction step and the finite number of iterations ($\sim\!3$) in the Kepler solver. 
On the hardware side, the only changes that would improve the runtime are an increase in clock speed or an increase in floating point execution units.
Neither of those changes seem likely given the current trend towards high parallelism and low precision aimed at machine learning applications.

Future work might focus on applying similar techniques to those presented in this paper to larger, arbitrary-$N$ simulations and to non-symplectic or hybrid-symplectic integrators that are capable of accurately resolving close encounters as well as hyperbolic orbits.
Some applications could also benefit from implementations of additional physics models such as tidal forces and moons, or from better models for general relativistic corrections.

\WHFastFiveTwelve is freely available in the \REBOUND integrator package at \url{https://rebound.hanno-rein.de}.
\REBOUND is compiled with \WHFastFiveTwelve by default on 64-bit x86 systems.
At runtime, \REBOUND checks for the relevant CPU flags and reports an error if the CPU does not support AVX512 instructions.
\WHFastFiveTwelve is also available in precompiled python wheels for Linux and Windows.

\begin{acknowledgments}
    We thank an anonymous referee for helpful comments which significantly improved the manuscript.
    We thank Daniel Tamayo and Dang Pham for helpful discussions and Dang Pham for help with testing \WHFastFiveTwelve on Windows.
    This research has been supported by the Natural Sciences and Engineering Research Council (NSERC) Discovery Grants RGPIN-2020-04513 and RGPIN-2026-05109.
    The authors did not use generative AI for code generation, or writing and editing this paper.
\end{acknowledgments}
\newpage

\appendix
\section{Kepler Solver for Very High Eccentricities}\label{app:highe}
\begin{figure*}[tb]
    \centering
\includegraphics[width=\textwidth]{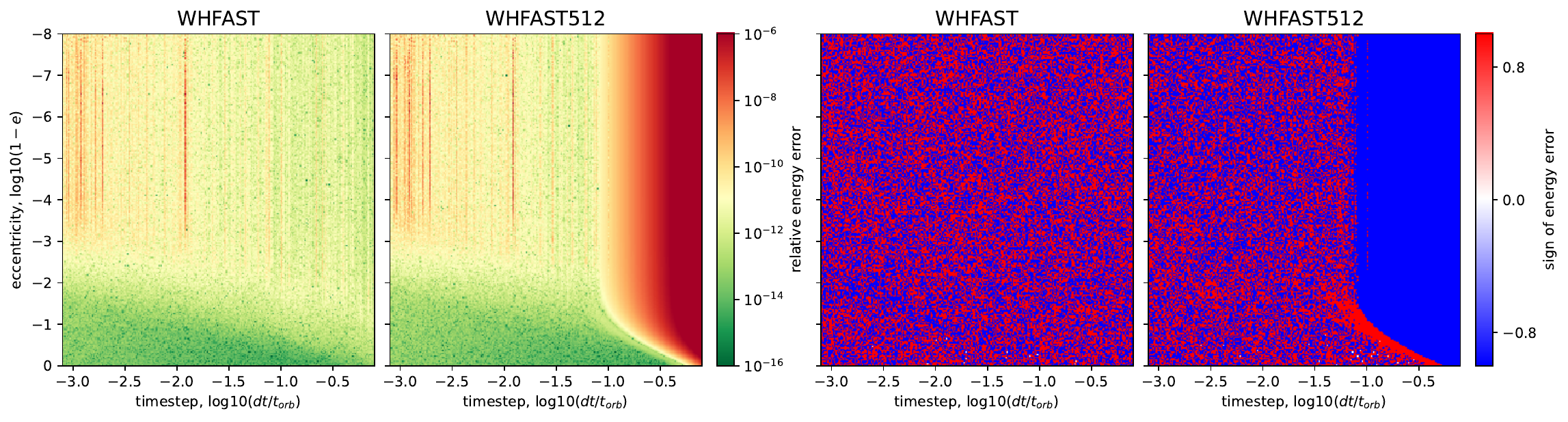} 
\caption{
    Comparison of the Kepler solvers in \WHFast and \WHFastFiveTwelve. We set $G=M=a=1$ and initialize one massless particle at pericenter. We integrate for 100~orbital periods. We sample $\log_{10}(dt/t_{\rm orb})$ from~$-3.1$~to~$-0.1$ and $\log_{10}(1-e)$ from~$0$~to~$-8$. The two left columns show the absolute relative energy error at the final time, and the two right columns show its sign.
    \label{fig:kepler_highe}}
\end{figure*}

We here compare the Kepler solvers in \WHFast and \WHFastFiveTwelve for very high eccentricities up to $e=1-10^{-8}$.
The results are plotted in Fig.~\ref{fig:kepler_highe}.
See \cite{ReinTamayo2015} for the precise setup used. Also compare our Fig.~\ref{fig:kepler_highe} to their Fig.~1.
For $e\leq0.99$ and $dt/t_{\rm orb}\leq0.05$, the maximum relative energy errors are $1.5\times10^{-11}$ for \WHFast and $9.1\times10^{-12}$ for \WHFastFiveTwelve. The energy errors are positive in 48\% and 50\% of the cases, respectively, showing that neither solver has a measurable bias in this region.
Note that \WHFastFiveTwelve achieves a slightly better accuracy than WHFast for eccentricities $\sim0.9-0.99$. We attribute this to our implementation using compensated summation for the Stiefel functions (see Sec.~\ref{sec:roundoff}).
For timesteps larger than approximately one tenth of the orbital period, the error of \WHFastFiveTwelve increases more quickly than that of \WHFast (see Sec.~\ref{sec:testkepler} for a discussion on this).

We point out that the parameter space tested here is in general not relevant for simulations of planetary systems if the goal is to determine a stable system's evolution or whether a system is unstable or not.
Further note that a standard Wisdom-Holman integrator such as \WHFast or \WHFastFiveTwelve might not be the right choice for systems with such high eccentricities because to resolve the shortest dynamical timescale in the problem (the pericenter timescale) one would need a prohibitively small timestep.

\section{Brute-force Search for an Efficient Kepler Solver}\label{app:bruteforce}
We tested different Kepler-solver sequences with one to three Newton or Halley refinement stages as well as Stiefel function truncation orders of 7, 11, 15, and 19.
Taken together, we tested 2336 different combinations.
For each combination, we measured the wall time and the errors in $e$, $a$, longitude of pericenter, and mean anomaly for eight Kepler problems with eccentricities ranging from 0.30 to 0.72 and a 6~day timestep.
We acknowledge that the fastest combination is ultimately depended on the specific planetary system benchmarked on. 
However, among the combinations that reached machine precision, several ones are almost equally fast.
The well performing combinations typically start with one or two Halley steps to quickly reach a moderately accurate solution but then use Newton steps to achieve the final convergence to machine precision.
Among the 2336 combinations, we select the fastest one.
It uses a 9-term Halley step, an 11-term Halley step, and then 19-term Newton iterations with a convergence check and a fallback to the bisection method if the Newton method does not converge within 4 step. 
The accuracy over a broader parameter space has then been verified, see Figs.~\ref{fig:kepler}, \ref{fig:kepler_long}, and \ref{fig:kepler_highe}.

\section{Roundoff-Bias Tests of Stiefel Functions}\label{app:roundoff_ablation}
We test the two roundoff corrections described in Section~\ref{sec:roundoff} with three versions of the final Stiefel function evaluation where we include both corrections, compensated summation only, and coefficient correction only. 
Using the notation from Algorithm~\ref{alg:stiefel}, the low-word recurrence is $D'=-zD+A(s_e-p_e)-C\delta_k$, where $A$ enables compensation of arithmetic roundoff and $C$ enables transport of the coefficient residual. We are thus testing $(A,C)=(1,1)$, $(1,0)$, and~$(0,1)$.

\begin{figure*}[tb]
    \centering
    \includegraphics[width=\textwidth]{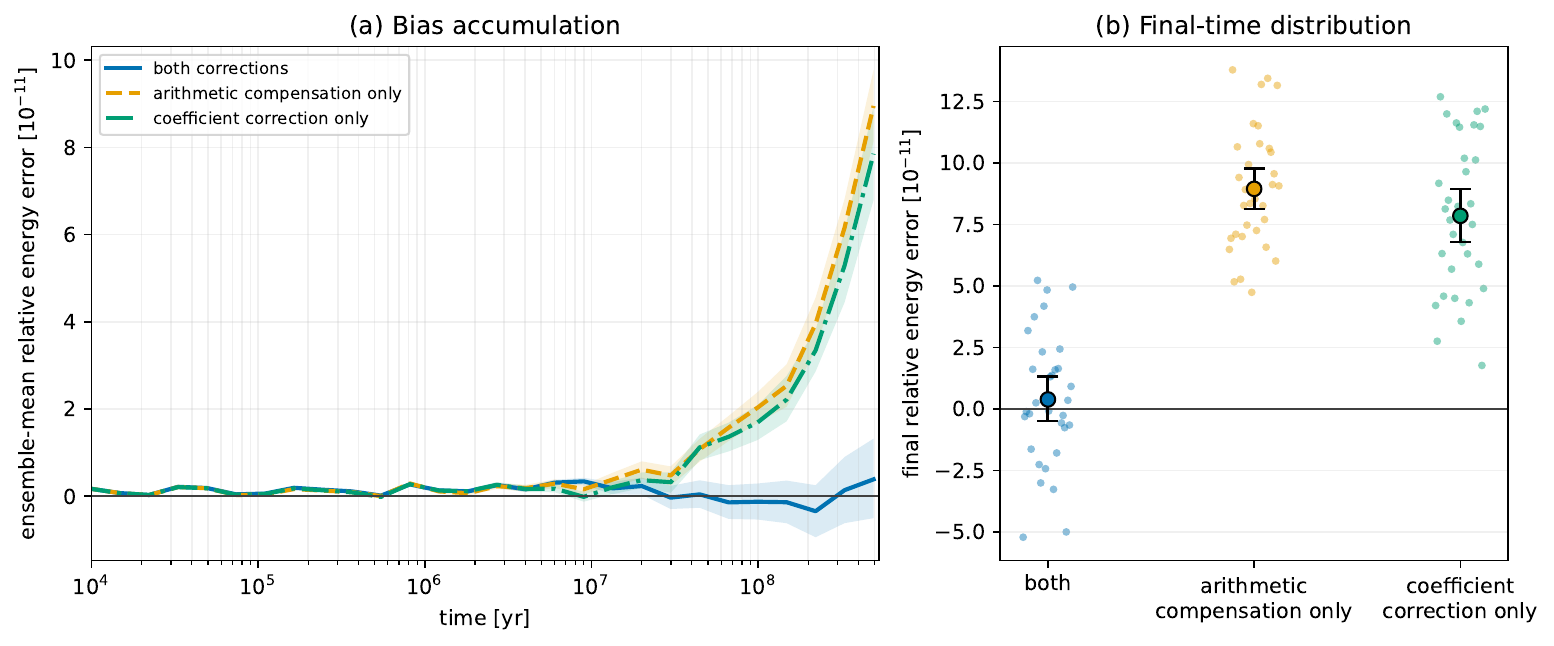}
    \caption{Roundoff-bias tests for sets of 32 Solar System integrations with both corrections, compensated summation only, and coefficient correction only. Left: ensemble-mean relative energy error as a function of time, shaded regions show two sigma confidence intervals. Right: final relative energy error for each simulation; small points show individual simulations, black points and error bars show the ensemble means and two sigma confidence intervals.}
    \label{fig:roundoff_ablation}
\end{figure*}

We integrate three sets of 32 simulations of the eight-planet Solar System with a 6-day timestep, general relativistic corrections, and an order-17 symplectic corrector for $3\times10^{10}$ timesteps. The initial positions and velocities of each simulation are perturbed at the $10^{-15}$ level. Figure~\ref{fig:roundoff_ablation} shows the ensemble-mean energy error as a function of time for each set in the left panel and the final result for each simulation in the right panel.

When both corrections are used the ensemble-mean energy error is $(0.4 \pm 1.2 )\cdot 10^{-11}$ at the final time, i.e. consistent with zero. 
With compensated summation only, the mean is $(8.6 \pm 0.8) \cdot 10^{-11}$; with coefficient correction only, it is $(7.9\pm 1.1 )\cdot10^{-11}$. 
Thus only using either one of the two corrections produces a positive bias, whereas using both does not have any bias over this interval.

\twocolumngrid

\bibliography{full}{}
\bibliographystyle{aasjournalv7}

\end{document}